\documentclass[11pt,a4paper]{article}

\usepackage[utf8]{inputenc}
\usepackage[T1]{fontenc}
\usepackage{lmodern}
\usepackage{microtype}
\usepackage{graphicx}
\usepackage{authblk}
\usepackage{orcidlink}
\usepackage{hyperref}
\usepackage[margin=1in]{geometry}

\title{\texttt{pySNOW}: a Python Suite for the NanO-World}

\author[1]{Sofia Zinzani\,\orcidlink{0009-0009-6961-7766}\thanks{Equal contribution. Corresponding author.}}
\author[1]{Gilberto Nardi\,\orcidlink{0009-0003-3276-8372}\thanks{Equal contribution. Corresponding author.}}
\author[1,2]{Giacomo Becatti\,\orcidlink{0009-0004-1686-7984}\thanks{Equal contribution.}}
\author[1]{Davide Alimonti\,\orcidlink{0009-0008-4530-1559}}
\author[1]{Letícia F. Basso\,\orcidlink{0009-0002-5108-3101}}
\author[3]{Kevin Rossi\,\orcidlink{0000-0001-8428-5127}}
\author[1]{Francesca Baletto\,\orcidlink{0000-0003-1650-0010}}

\affil[1]{Department of Physics, University of Milan, Via Celoria 16, Milan, 20133, Italy}
\affil[2]{Scientific Computing Center, Karlsruhe Institute of Technology, Hermann-von-Helmholtz-Platz 1, 76344 Eggenstein-Leopoldshafen}
\affil[3]{Materials Science and Engineering Department, Delft University of Technology, Delft, 2623CD, Netherlands}

\begin{document}

\maketitle


\section{Summary}


In computational materials science, numerical simulations are indispensable tools for revealing atomic-scale processes. For building blocks of the nanoworld, such as nanoparticles and nanoalloys, atomistic simulations provide detailed insight into their behaviour under diverse conditions. These simulations enable the study of formation and growth mechanisms, transport phenomena, chemophysical stability, and chemical reactions, including catalytic processes.

To unravel the complex structure--property relationships that characterise nanoobjects, it is crucial to develop robust and insightful representations of their atomistic structure at global and local scales. Such descriptions enhance our understanding of nanoparticle behaviour and play a key role in guiding their rational design in silico for targeted applications.

\section{Statement of need}

\texttt{pySNOW} -a Python Suite for the NanO-World- offers a complete Python workflow designed for the morphological characterisation of metallic nanoparticles: from input/output operations, through extended characterisation, up to connection with computational catalysis calculations and experimentally observable quantities. The main focus of pySNOW is the morphological characterisation of mono- and bi-metallic atomistic systems, e.g., from classical and ab initio molecular dynamics, Monte Carlo simulations, or eventually 3D tomography reconstruction. To this end, it provides the user with a suite of tools and routines along with a full list of tutorials. The only requirement is the coordinates or the trajectory of the nanosystem under study, in a simple, human-readable, standard .xyz format. From such inputs,

\texttt{pySNOW} allows the computation of several common descriptors, widely used in computational nanomaterials science \cite{baletto2019structural}, as well as more advanced observables that mimic the results of experimental techniques, such as SAXS spectra and chemical analyses.

Apart from morphological characterisation, it also enables catalytic investigation. \texttt{pySNOW} allows mapping the diversity of surface sites and adding one or more adsorbed molecules to them, providing input for further electronic structure calculations. Furthermore, it offers, within the CHE model \cite{Calle-Vallejo2014}, \cite{Rossi2020}, a microkinetic model for estimating specific current and mass activities during the oxygen reduction reaction (ORR) on Pt nanoparticles.

The motivation for developing the \texttt{pySNOW} package is the need for a unified and comprehensive toolkit for these analyses.  Thanks to its modular, functional and simple structure, it should be easily adopted by broad sections of both the computational and experimental materials science communities.


Apart from morphological characterisation, it also enables catalytic investigation. \texttt{pySNOW} allows mapping the diversity of surface sites and adding one or more asorbed structures to them, providing input for further electronic structure calculations. Furthermore, it offers, within the CHE model \cite{Calle-Vallejo2014}, \cite{Rossi2020}, a microkinetic model for estimating specific current and mass activities during the oxygen reduction reaction (ORR) on Pt nanoparticles.

The motivation for developing the \texttt{pySNOW} package is the need for a unified and comprehensive toolkit for these analyses.  Thanks to its modular, functional and simple structure, it should be easily adopted by broad sections of both the computational and experimental materials science communities.




\section{State of the field}

Several tools exist to analyse atomistic simulation results and MD trajectories. Here, we provide a quick overview of the main packages the authors are familiar with, along with their differences from pySNOW. MDTraj \cite{MDtraj} enables fast calculations of root-mean-square displacement (RMSD) and the extraction of common order parameters. Together with MDAnalysis \cite{MDAnalysis_gowers2016} \cite{MDAnalysis_Agrawal2011}, these packages are more oriented towards biomolecular systems. \texttt{pySCAL} \cite{pySCAL} is a Python module for the calculation of descriptors of local atomic environments, e.g. Steinhardt’s parameters, and with the possibility of adaptive cutoffs.

Freud \cite{freud:2020} is a capable and efficient Python/C++ tool that enables the computation of a wide range of quantities, including correlation functions and diffraction patterns.

ASE \cite{ase-paper}: is a general library capable of building and modifying structures and directly performing calculations. Along with these tools, it provides several post-processing tools, e.g. to obtain derived quantities such as phonon curves and vibration spectra, diffusion coefficients, or fitting equations of state.

While other packages provide characterisation tools, \texttt{pySNOW} covers a broad and extensive range of quantities. Being tailored for nanoparticles and bimetallic nanoalloys, it pays close attention to element-wise chemical analysis and also provides relevant experimentally accessible observables.



\section{Software design}

\texttt{pySNOW} has been developed with a few key design principles in mind: namely, (1) ease of use - to enable fast learning for a wide range of scientists with different backgrounds, (2) ease of modification, (3) independence from many complex packages, with the only required depdendencies being numpy and scipy, in order to maintain a high staibility over time and the reduce possible conflicts for future versions, (4) relying exclusively on open and freely available development environments and programming languages, and (5) integrability with other simulation and analysis codes.

We adopted a functional approach to implementing the software, avoiding the use of classes. This choice has been made to provide greater flexibility and easier access for users with a wide variety of programming skills.  We used a uniform style for function definitions and documented all relevant functions with complete docstrings to make them as user-friendly as possible.

\begin{figure}[htbp]
    \centering
    \includegraphics[width=\linewidth]{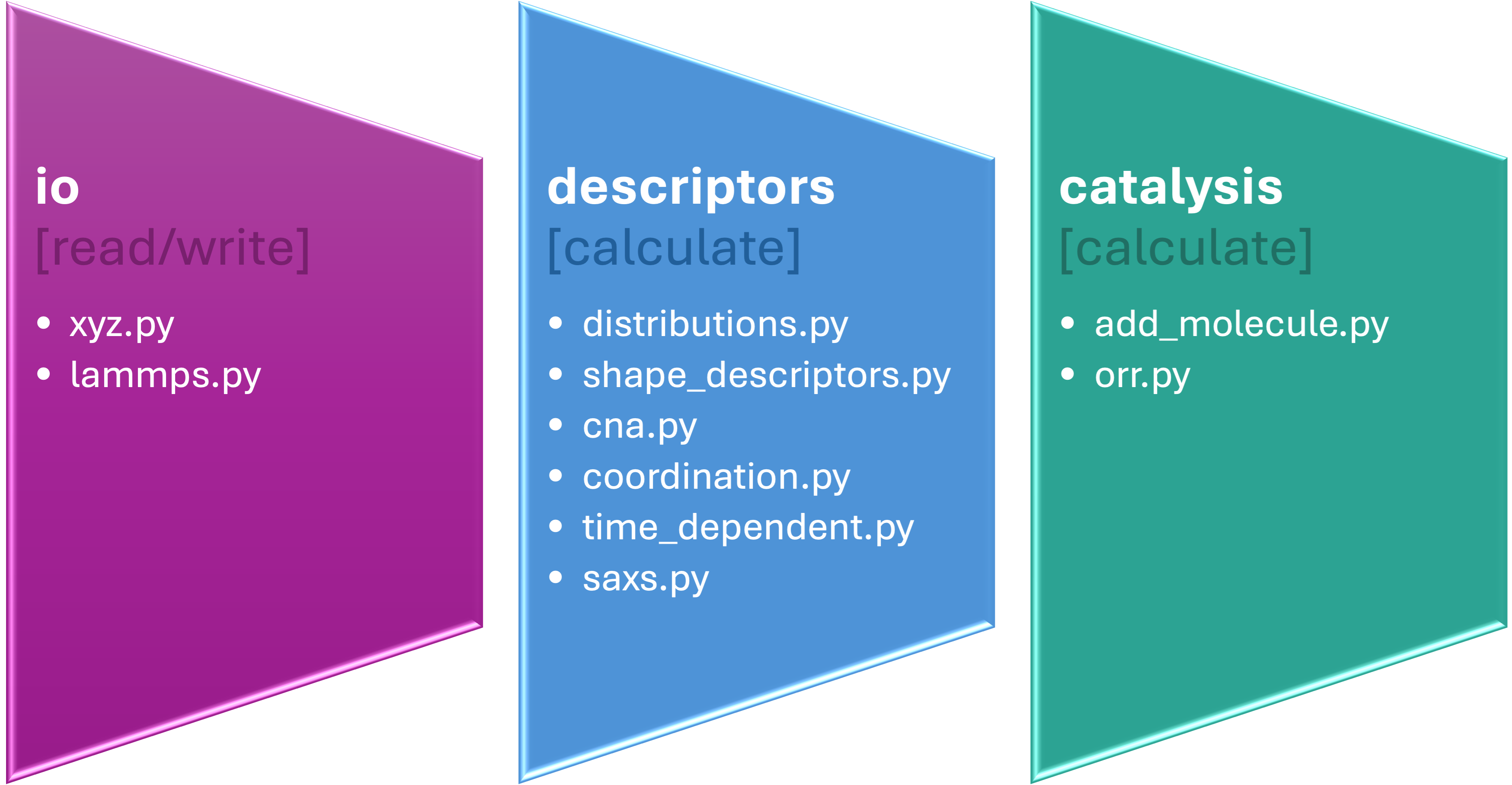}
    \caption{pySNOW workflow}
    \label{fig:pysnow-workflow}
\end{figure}

The code provides four main subpackages, as schematized in figure \ref{fig:pysnow-workflow}, including tools for reading/writing, analysing, and modifying atomistic configurations. These are:

\begin{itemize}

\item descriptors: the main pySNOW subpackage. It includes modules to compute several different descriptors. A comprehensive but non-exhaustive list includes: descriptors relevant to coordination (coordination numbers (CNs) and (strained) generalized CNs), common neighbour analysis (CNA) and CNA patterns, chemical element-wise atomic distributions (Pair Distance Distribution Function - PDDF, center of mass radial distribution function (RDF), pair radial density function, layer-by-layer density), nanoparticles’ general shape descriptors (including inertia tensor- and gyration tensor-based descriptors), chemical bond-based local atomic environment analysis, and Steinhardt parameters. It also includes a module to compute small-angle X-ray scattering (SAXS) spectra from both atomic coordinates and precomputed PDDFs.

\item io: the input/output subpackage. It includes modules for reading from and writing to file. Currently supported formats are .xyz, .lammps-data, and .lammps-dump. It supports both trajectory and single-frame files.

\item catalysis: a subpackage with functions oriented to the preparation and analysis of catalysis simulations. It includes the add\_molecule module, which allows adding a number of adsorbed molecules on the surface(s) of a nanoparticle, and the ORR module, which computes the current, the mass activity and the specific activity at an applied potential of a nanoparticle based on its atop generalised coordination number as in \cite{Rossi2020}.

\item misc: a subpackage with some utilities (the rototranslation module, to perform rototranslations of atomic coordinates) and tabulated constants (the constants module, which contains atomic masses and SAXS-related constants).

\end{itemize}

For some functions, a fast version is provided and recommended for situations with a large number of atoms to avoid time-calculation bottlenecks.

These modules are documented in the provided documentation. Together with the code documentation, we provide a list of Jupyter Notebook tutorials that cover all the relevant modules, their use in common use cases, tips and tricks.

Finally, along with the main modules, \texttt{pySNOW} also provides several unit tests that can be run automatically with the \texttt{pytest} package (which constitues an optional dependency of the package).


\section{Research Application (impact statement)}

\texttt{pySNOW} is applicable to morphological analysis  during the coalescence of nanoalloys \cite{Zinzani_Baletto:2026}; catalysis mapping \cite{Leti2025}; and catalytic performance \cite{Zinzani_Rossi:2025}.

\texttt{pySNOW} is taught at the Physics Department, Universita' degli Studi di Milano in a Master course, and it was used during undergraduate projects.

\section{AI usage disclosure}


While we consider our work original, we have used generative AI tools (OpenAI’s ChatGPT 5.3 and Anthropic’s Claude Opus 4.6) to compare solutions and check algorithms in parts of the code.

All AI-assisted outputs were reviewed, edited, and
validated by the human authors, who designed the overall code architecture and take full
responsibility for the accuracy and originality of the software and all submitted materials.



\section{Author contributions}

FB contributed to conceptualization, supervision, and financial support.

SZ, GB, and GN equally contributed to conceptualization, development, and documentation.

DA contributed to the SAXS part.

LFB contributed to GCN and molecule addition.

All authors conitrbuted to test the code, the tutorials, and contribute to the manuscript.

\section{Acknowledgements}

SZ, DA, GN, LFB acknowledge the Università degli Studi di Milano and the PhD programme at the Physics department. Furthermore, SZ thanks the financial support from ISC-SrL (D.M. 117/2023 PNRR).

DA thanks the CNR-ICCOM financial support for his PhD studentship (D.M. 630/2024 PNRR).

LFB thanks the CNR-Unimi collaboration for her PhD studentship and the PNRR (D.M. 118/TD/2023).

GB, FB, and SZ acknowledge financial support from the European Commission under the EIC Pathfinder project CHIRALFORCE, contract number 101046961. GN and FB thank MONSTER, a project of the Italian National Centre for HPC, Big Data and Quantum computing (ICSC, CUP B93C22000620006), funded by the European Union - NextGenerationEU, through PNNR, for its financial support.

FB and LFB thank the financial support from the FWO grant number G076625N on “Design of plasmonic photocatalysts for CO2 conversion based on multi-metallic clusters".

All authors thank the constructive discussion with R. M. Jones (Alan Turing Institute) and Mirko Vanzan (Unimi).

\section{Conflict of interest}

The authors declare no potential conflict of interests.

\section{References}

\bibliographystyle{unsrt}
\bibliography{paper}

\end{document}